# HETEROGENEOUS DISTRIBUTION OF [26]AL AT THE BIRTH OF THE SOLAR SYSTEM

Kentaro Makide[1], Kazuhide Nagashima[1], Alexander N. Krot[1*], Gary R. Huss[1], Fred J. Ciesla[2], Eric Hellebrand[3], Eric Gaidos[3], & Le Yang[2]

[1]Hawai'i Institute of Geophysics and Planetology, School of Ocean, Earth Science and Technology, University of Hawai'i at Mānoa, Honolulu, HI 96822, USA

*e-mail address of the corresponding author: sasha@higp.hawaii.edu

[2]Department of the Geophysical Sciences, University of Chicago, Chicago, IL 60637, USA

[3]Department of Geology and Geophysics, University of Hawai'i at Mānoa, Honolulu, HI 96822, USA

## ABSTRACT

It is believed that [26]Al, a short-lived ($t_{1/2}$ = 0.73 Ma) and now extinct radionuclide, was uniformly distributed in the nascent Solar System with the initial [26]Al/[27]Al ratio of ~5.2×10$^{-5}$, suggesting its external stellar origin. However, the stellar source of [26]Al and the manner in which it was injected into the solar system remain controversial: the [26]Al could have been produced by an asymptotic giant branch star, a supernova, or a Wolf-Rayet star and injected either into the protosolar molecular cloud or protoplanetary disk. Corundum ($Al_2O_3$) is thermodynamically predicted to be the first condensate from a cooling gas of solar composition. Here we show that micron-sized corundum condensates from [16]O-rich gas ($\Delta^{17}O \sim -25‰$) of solar composition recorded heterogeneous distribution of [26]Al at the birth of the solar system: the inferred initial [26]Al/[27]Al ratio ranges from ~6.5×10$^{-5}$ to <2×10$^{-6}$; ~50% of the corundum grains measured are [26]Al-poor.

Other $^{26}$Al-poor, $^{16}$O-rich refractory objects include grossite ($CaAl_4O_7$)- and hibonite ($CaAl_{12}O_{19}$)-rich calcium-aluminum-rich inclusions (CAIs) in CH chondrites, platy hibonite crystals in CM chondrites, and FUN (fractionation and unidentified nuclear isotopic anomalies) CAIs in CV, CO, and CR chondrites. Considering the apparently early and short duration (<0.3 Ma) of condensation of refractory $^{16}$O-rich solids in the solar system, we infer that $^{26}$Al was injected into the collapsing protosolar molecular cloud and later homogenized in the protoplanetary disk. The apparent lack of correlation between $^{26}$Al abundance and O-isotope compositions of corundum grains put important constraints on the stellar source of $^{26}$Al in the solar system.

# 1. INTRODUCTION

$^{26}$Al (decays to $^{26}$Mg with a half-life, $t_{1/2} = 0.73$ Ma), $^{10}$Be ($t_{1/2} = 1.5$ Ma), $^{36}$Cl ($t_{1/2} = 0.3$ Ma), $^{41}$Ca ($t_{1/2} = 0.1$ Ma), $^{53}$Mn ($t_{1/2} = 3.7$ Ma), and $^{60}$Fe ($t_{1/2} = 2.3$ Ma) are short-lived radionuclides that were present in the nascent Solar System (McKeegan & Davis 2003). The recently inferred uniform distribution of $^{26}$Al, at least in the inner part of the solar system (Jacobsen et al. 2008, Villeneuve et al. 2009, Davis et al. 2010), makes $^{26}$Al-$^{26}$Mg system one of the most important relative chronometers of the earliest solar system processes, such as evaporation, condensation, and melting of solids in the protoplanetary disk (McKeegan & Davis 2003, Kita et al. 2005, Krot et al. 2009). The maximum amount of $^{26}$Al that could have been produced by solar energetic particle irradiation is insufficient to explain the canonical $^{26}$Al/$^{27}$Al ratio of ~5.2×10$^{-5}$ in the

inner part of the solar system and suggests a stellar, nucleosynthetic origin of $^{26}$Al (Duprat & Tatischeff 2007). The stellar source of $^{26}$Al and the way by which it was injected into the solar system, however, remain controversial. The freshly synthesized $^{26}$Al from a neighboring asymptotic giant branch (AGB) star (Wasserburg et al. 1994), a supernova (SN) (Boss et al. 2008), or a Wolf-Rayet star (Arnould et al. 2006, Gaidos et al. 2009, Tatischeff et al. 2010) could have been injected either into the protosolar molecular cloud (Boss et al. 2008) or protoplanetary disk (Ouellette et al. 2007).

An AGB source of $^{26}$Al in the solar system is highly unlikely, because AGB stars are not associated with star-forming regions and the probability of an encounter between an AGB star and a star-forming molecular cloud is very low, $\sim 2.5 \times 10^{-6}$ (Huss et al. 2009). Although massive SN progenitors are commonly associated with star-forming regions (Lada & Lada 2003), the probability of injection of $^{26}$Al into the protosolar molecular cloud or protoplanetary disk from a nearby SN is also estimated to be low, $\leq 3 \times 10^{-3}$ (Williams & Gaidos 2007, Gounelle & Meibom 2008). To overcome this problem, it has been recently suggested that the Sun was born in a stellar cluster of second generation that was self-enriched in short-lived radionuclides ($^{60}$Fe and possibly $^{26}$Al) by a single or multiple 10–15 solar mass ($M_{\odot}$) SN several Ma before our solar system formed (Gounelle et al. 2009). Finally, it was proposed that $^{26}$Al could have been injected into the protosolar molecular cloud by wind from a Wolf-Rayet star (Arnould et al. 2006, Gaidos et al. 2009, Tatischeff et al. 2010).

Calcium-aluminum-rich inclusions (CAIs) are the oldest solar system solids dated (Amelin et al. 2002, Bouvier & Wadhwa 2010) and believed to have formed when the Sun was a class 0 or class I protostar (e.g., Krot et al. 2009; Tscharnuter et al. 2009) though Ciesla (2010) argued that most CAIs found in meteorites formed at the time that the solar nebula stopped accreting significant amounts of material from the protosolar parent cloud, and thus corresponds roughly to the class I − class II transition. The majority of CAIs in primitive (unmetamorphosed) chondrites have the initial $^{26}$Al/$^{27}$Al ratio (($^{26}$Al/$^{27}$Al)$_0$) of ~(4−5)×10$^{-5}$ (MacPherson et al. 1995, Jacobsen et al. 2008, Makide et al. 2009a). CAIs with low ($^{26}$Al/$^{27}$Al)$_0$, <<5×10$^{-6}$, are very rare (Makide et al. 2009a, Sahijpal & Goswami 1998, Liu et al. 2009), except in CH carbonaceous chondrites where they are dominant (Krot et al. 2008). The $^{26}$Al-poor CAIs are interpreted either as a result of early formation, prior to injection and homogenization of $^{26}$Al in the solar system (Makide et al. 2009a, Sahijpal & Goswami 1998, Liu et al. 2009, Krot et al. 2008) or as a result of preferential loss of $^{26}$Al carrier from CAI precursors by sublimation (Thrane et al. 2008). The late-stage formation of $^{26}$Al-poor CAIs, after decay of $^{26}$Al, can not be also entirely excluded, because CAIs are known to have experienced multistage thermal processing in the solar nebula and on chondrite parent asteroids (Krot et al. 2005) that could have erased possible evidence for radiogenic $^{26}$Mg (MacPherson et al. 1995).

Corundum ($Al_2O_3$) is thermodynamically predicted to be the first condensate from a cooling gas of solar composition (Ebel & Grossman 2000). Therefore, corundum condensates can potentially provide important constraints on the origin of $^{26}$Al and degree of its heterogeneity in the nascent solar system. The pioneering work on solar corundum grains was done by Virag et al. (1991) who studied 26 individual corundum grains 3−20 μm in size from the CM carbonaceous chondrite Murchison. On the basis of the oxygen and magnesium isotopic compositions, the grains were divided into three groups. Group 1 (n = 17) and group 2 (n = 5) grains show $^{26}$Mg excesses corresponding to ($^{26}$Al/$^{27}$Al)$_0$ of 5×10$^{-5}$ and 5×10$^{-6}$, respectively; group 3 grains (n = 4) show no resolvable $^{26}$Mg excesses, ($^{26}$Al/$^{27}$Al)0 < 3×10$^{-7}$.

To understand the origin and distribution of $^{26}$Al in the early solar system, we measured $^{26}$Al-$^{26}$Mg isotope systematics and oxygen isotopic compositions of micron (μm)-sized corundum grains from acid-resistant residues of unequilibrated ordinary (H and LL of low petrologic type, ≤3.15) and unmetamorphosed carbonaceous chondrites (CI1, CM2, CR2, and CO3.0) using the University of Hawai'i Cameca ims-1280 secondary ion mass-spectrometer. Oxygen isotopes were used to distinguish solar from presolar corundum grains; the latter are typically characterized by isotopically anomalous O-isotope compositions (e.g., Nittler 2003) compared to the solar and terrestrial values, $\Delta^{17}$O ∼ −25‰ and 0‰, respectively (Makide et al. 2009a, McKeegan et al. 2010), where $\Delta^{17}$O = $\delta^{17}$O − 0.52×$\delta^{18}$O, $\delta^{17,18}$O = (($^{17,18}$O/$^{16}$O)$_{sample}$)/($^{17,18}$O/$^{16}$O)$_{SMOW}$ − 1) × 1000, and SMOW is Standard Mean Ocean Water.

## 2. ANALYTICAL PROCEDURES

### 2.1. Sample preparation

Acid-resistant residues of ordinary and carbonaceous chondrites used in this study were prepared by Huss and Lewis (1995). The residues consist of silicon carbide (SiC), spinel (MgAl$_2$O$_4$), hibonite (CaAl$_{12}$O$_{19}$), and corundum and range in size from 0.5 to 5 μm. The residues were diluted by a mixture of 90% isopropanol and 10% water and mixed in an ultrasonic bath. About 0.2 μl of the mixture was siphoned using a micro-aspirator and dispersed onto the gold substrate under a stereomicroscope. More than 100 grains were dispersed on each gold substrate. Prior to the dispersion, the gold substrate has been examined for possible contamination using the University of Hawai'i (UH) JEOL JXA-8500F field emission electron microprobe (FE-EPMA) equipped with a cathodoluminescence detector (CL, Gatan Mini-CL) and an energy dispersive spectrometer (EDS, Thermo UltraDry). Several relatively large (> 5 μm) corundum grains embedded in the substrates were identified and avoided during subsequent oxygen and magnesium isotopic measurements. To locate the chondritic corundum grains comprising less than 1% of the dispersed grains dominated by spinel, we used CL and EDS of the FE-EPMA. Among the dispersed grains, only corundum and silicon carbide show intense CL; the latter is very rare and can be easily distinguished by EDS. The corundum grains identified were photographed in secondary and backscattered electrons.

## 2.2. Oxygen-isotope measurements

Oxygen isotopic compositions of individual corundum grains were measured with the UH Cameca ims-1280 secondary ion mass-spectrometer (SIMS or ion microprobe). The details of the analytical technique used are described by Makide et al. (2009b).

## 2.3. Magnesium-isotope measurements

Magnesium isotopic measurements with UH Cameca ims-1280 were conducted on the corundum grains which have been previously analyzed for oxygen isotopes. An $^{16}O^{-}$ primary beam, either defocused (~30 μm) or focused (~3 μm), was used to generate secondary $Mg^{+}$ and $Al^{+}$ ions. A field aperture of 1000×1000 μm$^2$ corresponding to ~7 μm on the sample was used to minimize contribution of Mg signals from the substrate and other dispersed grains surrounding the grain of interest. The contribution of Mg signal from the substrate was estimated to be <1% of Mg from the sample. The mass resolving power was set to ~3500−3800, sufficient to separate interfering hydrides and $^{48}Ca^{++}$. Secondary ions of $^{24}Mg^{+}$, $^{25}Mg^{+}$, and $^{26}Mg^{+}$ were measured simultaneously using the monocollector EM and two multicollector EMs; subsequently $^{27}Al^{+}$ was measured with the monocollector EM by peak-jumping. The primary beam current was adjusted so that the $^{27}Al^{+}$ count rate did not exceed 4.5×10$^5$ cps. Acquisition times for Mg isotopes and $^{27}Al$ were 15 sec and 1 sec, respectively. Each measurement consisted of 20 cycles. Excess of ra-diogenic $^{26}$Mg ($\delta^{26}$Mg*) was calculated using the following equation: $\delta^{26}$Mg* = $\Delta^{26}$Mg −
2×$\Delta^{25}$Mg, where $\Delta^{25,26}$Mg = (($^{25,26}$Mg/$^{24}$Mg)$_{sample}$)/($^{25,26}$Mg/$^{24}$Mg)$_{ref}$ − 1) × 1000 and reference
Mg isotope ratios are from Catanzaro et al. (1966). The typical reproducibility of standard meas-
urements was ~10‰ (2σ) in $\Delta^{25}$Mg and $\Delta^{26}$Mg. To correct for instrumental mass fractionation
effect and estimate relative sensitivity factor for $^{27}$Al/$^{24}$Mg ratio, we used a Montana sapphire
standard (NMNH 126321 from Smithsonian Institution kindly provided by Glenn J. MacPher-
son). The instrumental fractionation correction was made based on the measurements of μm-
sized sapphire grains dispersed on the gold substrate. To correct the $^{27}$Al$^+$/$^{24}$Mg$^+$ ratios measured
by SIMS in μm-sized corundum grains, we measured Mg concentration in the Montana sapphire
by the UH FE-EPMA and multicollector inductively-coupled plasma mass-spectrometer (MC-
ICPMS) by Edward D. Young (UCLA). To measure trace amount of Mg in the sapphire by FE-
EPMA, we used defocused (10 μm) electron beam at ~300 nA current and 15 keV accelerating
voltage. Madagascar hibonite was used as a standard. The $^{27}$Al/$^{24}$Mg ratios in the sapphire ob-
tained by FE-EPMA and LA MC-ICP-MS are 6084±498 (2σ, 45 spots) and 5830±510
(2σ, 4 spots) are the same within the uncertainty of measurements and show no evidence for het-
erogeneity.

## 3. RESULTS AND DISCUSSION

### 3.1. Oxygen- and magnesium-isotope compositions of micron-sized corundum grains

Oxygen isotopic compositions of μm-sized corundum grains from acid-resistant residues of
unequilibrated ordinary chondrites (Semarkona (LL3.0), Bishunpur (LL3.1), Roosevelt County
(RC) 075 (H3.1)) and unmetamorphosed carbonaceous chondrites (Orgueil (CI1), Murray
(CM2), Renazzo (CR2), and Allan Hills (ALH) A77307 (CO3.0)) together with compositions of the mineralogically pristine CAIs from CR2 carbonaceous chondrites (Makide et al. 2009a) and the solar wind returned by the Genesis spacecraft (McKeegan et al. 2010) are listed in on-line Table 1 and plotted in Figure 1. All but three corundum grains have $^{16}$O-rich compositions similar to those of CR CAIs and solar wind, consistent with a condensation origin from an $^{16}$O-rich gas of solar composition. Three corundum grains, not shown in Figure 1, have highly anomalous O-isotope compositions, and are probably presolar in origin.

The solar corundum grains have been subsequently measured for magnesium-isotope compositions and $^{27}$Al/$^{24}$Mg ratio; the results are listed in on-line Table 1 and plotted in Figure 2. The grains show large variations of the inferred initial $^{26}$Al/$^{27}$Al ratio (($^{26}$Al/$^{27}$Al)$_0$): 52% of grains have no resolvable $^{26}$Mg*: an upper limit on ($^{26}$Al/$^{27}$Al)$_0$ is $2\times10^{-6}$; 40% of grains have high ($^{26}$Al/$^{27}$Al)$_0$, $(3.0–6.5)\times10^{-5}$; 8% of grains have intermediate values of ($^{26}$Al/$^{27}$Al)$_0$, $(1–2)\times10^{-5}$.

The coexistence of the $^{26}$Al-rich and $^{26}$Al-poor corundum grains in the same primitive meteorite preclude late-stage (after decay of $^{26}$Al) resetting of the $^{26}$Al-$^{26}$Mg systematics of the $^{26}$Al-poor corundum grains during thermal metamorphism on the host chondrite parent asteroids. Late-stage resetting in the solar nebula during thermal processing associated with chondrule formation can be also excluded: although chondrule formation appears to have lasted for 2–3 Ma (e.g., Amelin et al. 2002, Kita et al. 2005, Kurahashi et al. 2008, Krot et al. 2009), it occurred in an $^{16}$O-poor ($\Delta^{17}$O $> -10$‰) nebular reservoir (e.g., Krot et al. 2006 and references therein) when $^{26}$Al was still alive. We infer that $^{16}$O-rich corundum grains with low initial $^{26}$Al/$^{27}$Al ratios never contained high abundance of $^{26}$Al, i.e., the lack of $^{26}$Mg excess in ~50% of $^{16}$O-rich corundum grains is their primary characteristic.

*3.2. Multiple generations of CAIs and refractory grains and their distribution in the protoplanetary disk*

High abundance (>50%) of $^{26}$Al-poor, $^{16}$O-rich CAIs has been previously reported in metal-rich CH carbonaceous chondrites (Krot et al. 2006). Similar to corundum grains, the CH CAIs are among the smallest (< 20 μm in apparent diameter) and the most refractory (rich in grossite, $CaAl_4O_7$ and hibonite, $CaAl_{12}O_{19}$) objects known (Krot et al. 2002 and references therein). The $^{26}$Al heterogeneity and high abundance of $^{26}$Al-poor refractory objects (corundum grains and CH CAIs) in a fine fraction (<20 μm) of solar dust can be explained in the context of the formation setting of refractory objects (MacPherson et al. 1995, Krot et al. 2009) and theoretical modeling of their radial transport in the protoplanetary disk (Boss 2008, Ciesla 2009, Ciesla 2010, Yang & Ciesla 2010).

It is commonly inferred that refractory objects formed in region(s) with high ambient temperature (at or above condensation temperature of forsterite ($Mg_2SiO_4$), ~ 1350 K at total pressure of $<10^{-4}$ bar), most likely within 1–2 astronomical units from the Sun (Krot et al. 2009 and references therein), and were subsequently transported outward to the accretion regions of chondrite parent asteroids and cometary nuclei (Gounelle & Meibom 2007, Boss 2008, Ciesla 2009, Ciesla 2010). The oldest $^{207}$Pb-$^{206}$Pb absolute ages of CV CAIs (Amelin et al. 2002, Bouvier & Wadhwa 2010), a narrow range of their initial $^{26}$Al/$^{27}$Al ratios (Jacobsen et al. 2008), corresponding to a range of crystallization ages of <0.1 Ma, and the high ambient temperature in the CAI-forming region(s), are all consistent with an early and brief duration of CAI formation, during a period of high mass-accretion rate of dust and gas to the proto-Sun (~$10^{-6}$ $M_\odot$/year) (Krot et al. 2009, Ciesla 2009). We suggest that the $^{26}$Al-poor and $^{26}$Al-rich $^{16}$O-rich corundum grains represent different generations of refractory objects formed during a brief epoch of CAI formation, and, therefore, reflect heterogeneous distribution of $^{26}$Al at the birth of the solar system. The interpretation is consistent with a bi-modal distribution of the initial $^{26}$Al/$^{27}$Al ratios (~$5\times10^{-5}$ and

<<5×10$^{-6}$) among $^{16}$O-rich, small, and very refractory CAIs in primitive chondrites (Simon et al. 2002, Krot et al. 2008, Liu et al. 2009).

To understand the distribution of refractory objects with the different $(^{26}Al/^{27}Al)_0$ in the early solar system, we have modeled the evolution of a viscous disk and redistribution of refractory objects formed in the high-temperature (above 1400 K) region within it during the first Ma of its evolution (Fig. 3). We consider both the mass and angular momentum transport in the disk as well as the addition of mass to both the Sun and the disk due to infall from the parent molecular cloud (Yang & Ciesla 2010, Hueso & Guillot 2005). The parent cloud is assumed to be 1 M$_\odot$, spherical, at a temperature of 15 K, rotating with a solid angular velocity of $\Omega_c = 10^{-14}$ s$^{-1}$, and falls inward with a mass accretion rate of ~3×10$^{-6}$ M$_\odot$/yr; the disk viscosity is parameterized with a turbulent viscosity coefficient of $\alpha = 10^{-3}$. The accreted mass falls onto the disk at locations where the specific angular momentum in the disk equals that in the parent cloud (Yang & Ciesla 2010, Hueso & Guillot 2005). We track populations of fine refractory objects as they get redistributed throughout the disk by viscous expansion, dividing them into batches depending on the timing of when they are last exposed to the high-temperature region (Figs. 3a,b)

We find that the mass of the disk reaches its maximum by 0.3 Ma of its evolution (Fig. 3a), and that the most dominant population of refractory objects in a 1 Ma-old disk are those that formed immediately before and after infall from the parent molecular cloud ceased (Fig. 3b). These results constrain the time when $^{26}$Al was introduced into the solar system: it probably occurred contemporaneously with the collapse of the molecular cloud; injection into the disk (Ouellette et al. 2007) is less likely. They also indicate that if $^{26}$Al was heterogeneously distributed in the parent molecular cloud, mixing in the disk could not homogenize it until the infall had ceased. Thus, refractory objects formed prior to the end of collapse would not record the average solar nebula $^{26}$Al/$^{27}$Al ratio, but rather the ratio that existed in the hot nebular region at the time of their formation. This would allow refractory objects with different initial $^{26}$Al/$^{27}$Al ratios ($^{26}$Al-poor and $^{26}$Al-rich) to form, and, as we have shown, then be preserved in the disk. It also implies that the initial $^{26}$Al/$^{27}$Al ratio of $(5.23\pm0.13)\times10^{-5}$ recorded by CV CAIs (Jacobsen et al. 2008, Davis et al. 2010) may not necessarily represent the average $^{26}$Al/$^{27}$Al ratio in the solar system. For example, based on the high-precision whole-rock Mg-isotope isotope measurements of CI carbonaceous chondrites using multicollector inductively coupled plasma mass-spectrometer (MC-ICPMS), Larsen et al. (2010) concluded that the initial $^{26}$Al/$^{27}$Al ratio in the protoplanetary disk was much lower ($\sim 2.8\times10^{-5}$) than the canonical $^{26}$Al/$^{27}$Al ratio recorded by the CV CAIs. Additional work is needed to confirm this result.

Our modeling indicates that a significant population of refractory objects formed prior to the cessation of infall survives in the disk (Fig. 3b). The survival time of dust particles (up $\sim 1$ meter in size) in the disk is inversely proportional to their sizes: fine (5 $\mu$m) dust is well-coupled to the nebular gas and less affected by head wind than the coarse ($\sim 1$ mm) dust (Ciesla 2010). These observations may explain the high proportion of $^{26}$Al-poor refractory objects in a fine fraction of solar dust as we have observed in corundum grains and CH CAIs, as well as the presence of an $^{26}$Al-poor $^{16}$O-rich CAI in comet Wild2 (Matzel et al. 2010).

### 3.3. Injection of $^{26}$Al from a massive (> 30 $M_\odot$) star

The heterogeneous distribution of $^{26}$Al at the birth of the solar system may preclude inheritance of $^{26}$Al from a self-enriched molecular cloud (Gounelle et al. 20009), which is expected to have a uniform distribution of $^{26}$Al. Instead, it indicates injection of $^{26}$Al either into the collapsing molecular cloud or protoplanetary disk. The similar oxygen isotopic compositions of the $^{26}$Al-rich and $^{26}$Al-poor corundum grains and the Sun suggest injection of $^{26}$Al by a massive (>

30 M$_\odot$ star, either a SN or a Wolf-Rayet star, because injection of $^{26}$Al by less massive stars are expected to produce significant change in oxygen-isotope composition of the solar system (Gounelle & Meibom 2007, Krot et al. 2008, Ellinger et al. 2010).

To distinguish between a SN and a Wolf-Rayet sources of $^{26}$Al, measurements of other isotopes (e.g., $^{60}$Fe-$^{60}$Ni isotope systematics) of $^{26}$Al-rich and $^{26}$Al-poor $^{16}$O-rich CAIs or refractory grains are required: in contrast to supernovae, the Wolf-Rayet stars do not produce $^{60}$Fe (Sahijpal & Soni 2006). Therefore, if freshly-synthesized $^{26}$Al was injected by a Wolf-Rayet star (Gaidos et al. 2009, Tatischeff et al. 2010) and $^{60}$Fe was inherited from a molecular cloud of an earlier generation (Gounelle et al. 2009), the $^{26}$Al and $^{60}$Fe would be decoupled: i.e., the $^{26}$Al-rich and $^{26}$Al-poor CAIs will have similar initial abundance of $^{60}$Fe. In contrast, if $^{26}$Al and $^{60}$Fe were injected by the same SN, the $^{26}$Al-poor CAIs would have no or lower initial abundance of $^{60}$Fe than the $^{26}$Al-rich CAIs.

The initial abundance of $^{60}$Fe in the solar system remains controversial. Based on the Fe-Ni isotope systematics of chondrules in unequilibrated ordinary chondrites and differentiated meteorites (angrites and eucrites), the inferred initial $^{60}$Fe/$^{56}$Fe ratios range from $(2-5)\times10^{-7}$ (Telus et al. 2011) to $<1\times10^{-8}$ (Tang & Dauphas 2011, Spivak-Birndorf et al. 2011). Due to the presence of nucleosynthetic isotopic anomalies in Ni and low Fe/Ni ratios in CAIs (2–20), the unambiguous detection of radiogenic $^{60}$Ni in whole-rock CAIs is difficult (Quitté et al. 2007) and may require high-precision Fe-Ni isotope measurements of mineral separates of CAIs not to obtain an internal Fe-Ni isochron which have not been done yet.

We thank M. Bizzaro, N. Dauphas, M. Gounelle, I. D. Hutcheon, T. R. Ireland, K. Liffman, L. Nittler, J. R. Lyons, K. D. McKeegan, Q.-Z. Yin, and H. Yurimoto for fruitful conversations. Constructive review by Dr. Roberto Gallino is highly appreciated. We thank Dr. E. D. Young (UCLA) for help with LA-ICPMS measurements of saphire standard kindly provided by Dr. G. J. MacPherson (Smithsonian Institution). This research has received support from NASA grants NNX08AH91G and NNX07AI81G (ANK). This is Hawai'i Institute of Geophysics and Planetology publication No. XXX and School of Ocean and Earth Science and Technology publication No. YYY.

**FIGURE CAPTIONS**

**Figure 1.** $\Delta^{17}O$ values of μm-sized corundum grains, mineralogically pristine CAIs from CR carbonaceous chondrites (Makide et al. 2009a), and solar wind returned by the Genesis space-

craft (McKeegan et al. 2010). Errors are 2σ. The $^{16}$O-rich composition of corundum grains similar to those of CR CAIs and solar wind is consistent with a condensation origin from a gas of solar composition. Petrologic types of ordinary chondrites are from (Grossman & Brearley 2005).

**Figure 2. (a)** Aluminum-magnesium evolutionary isotope diagram for the μm-sized corundum grains #02-01 and #01-10 from the CI carbonaceous chondrite Orgueil. The corundum grain #02-01 has large excesses of $^{26}$Mg corresponding to the initial $^{26}$Al/$^{27}$Al ratio of $(4.7\pm0.9)\times10^{-5}$. No resolvable excess of $^{26}$Mg is found in the corundum grain #01-10; an upper limit on the initial $^{26}$Al/$^{27}$Al $((^{26}$Al/$^{27}$Al$)_0)$ ratio is $4.8\times10^{-7}$. Errors are 2σ. **(b)** $\Delta^{17}$O values *vs.* $(^{26}$Al/$^{27}$Al$)_0$ in μm-sized corundum grains from unequilibrated ordinary (UOCs) and primitive carbonaceous chondrites (CCs). Errors are 2σ. There is a bi-modal distribution of $(^{26}$Al/$^{27}$Al$)_0$: about 52% of grains have no detectable excess of $^{26}$Mg (an upper limit on $(^{26}$Al/$^{27}$Al$)_0$ is $2\times10^{-6}$); about 40% of grains have large excesses of $^{26}$Mg corresponding to $(^{26}$Al/$^{27}$Al$)_0$ from $\sim3.0\times10^{-5}$ to $\sim6.5\times10^{-5}$; 8% of grains have intermediate values. There is no correlation between O- and $^{26}$Al-$^{26}$Mg isotope systematics of corundum grains. In contrast to the corundum grains, the $(^{26}$Al/$^{27}$Al$)_0$ in whole-rock CV CAIs shows small variations, $(5.23\pm0.13)\times10^{-5}$. **(c)** The distribution of $(^{26}$Al/$^{27}$Al$)_0$ in solar corundum grains.

**Figure 3. (a)** Masses of the solar mass star, disk, and star + disk as a function of time. The mass of the disk grows to about 0.3 $M_\odot$ due to infalling of molecular cloud material, then decreases as the material is transported inward and accretes into the central star. **(b)** Fraction of dust processed at high temperature (> 1400 K) and survived in a 1 Myr-old disk, plotted in different batches (numbered from 1 to 12) according to when they last were exposed to those high temperatures.

On-line Table 1. Oxygen isotopic composition (in ‰) and the inferred $(^{26}Al/^{27}Al)_0$ ratio from individual μm-sized corundum grain in unequilibrated ordinary and carbonaceous chondrites

| grain | $\delta^{18}O$ | $2\sigma$ | $\delta^{17}O$ | $2\sigma$ | $\Delta^{17}O$ | $2\sigma$ | $(^{26}Al/^{27}Al)_0 \times 10^{-5}$ | $2\sigma$ |
|---|---|---|---|---|---|---|---|---|
| *Carbonaceous chondrites* | | | | | | | | |
| ALHA 03-03 | -35.9 | 2.3 | -43.4 | 4.5 | -24.7 | 4.7 | N.D. | |

…

Sem = Semarkona; Bis = Bishunpur; RC075 = Roosevelt County 075; ALHA = Alan Hills A77307; Murr = Murray; Org = Orgueil; Ren = Renazzo; *P = presolar corundum grains; *U = isotopically unusual corundum grains, possibly an analytical artifact. The *P and *U grains are not plotted in Figure 1 and not discussed in the text. N.D. = not determined due to either the presence of Mg-rich grains attached to corundum grains or loss of grains during oxygen isotopic measurements.

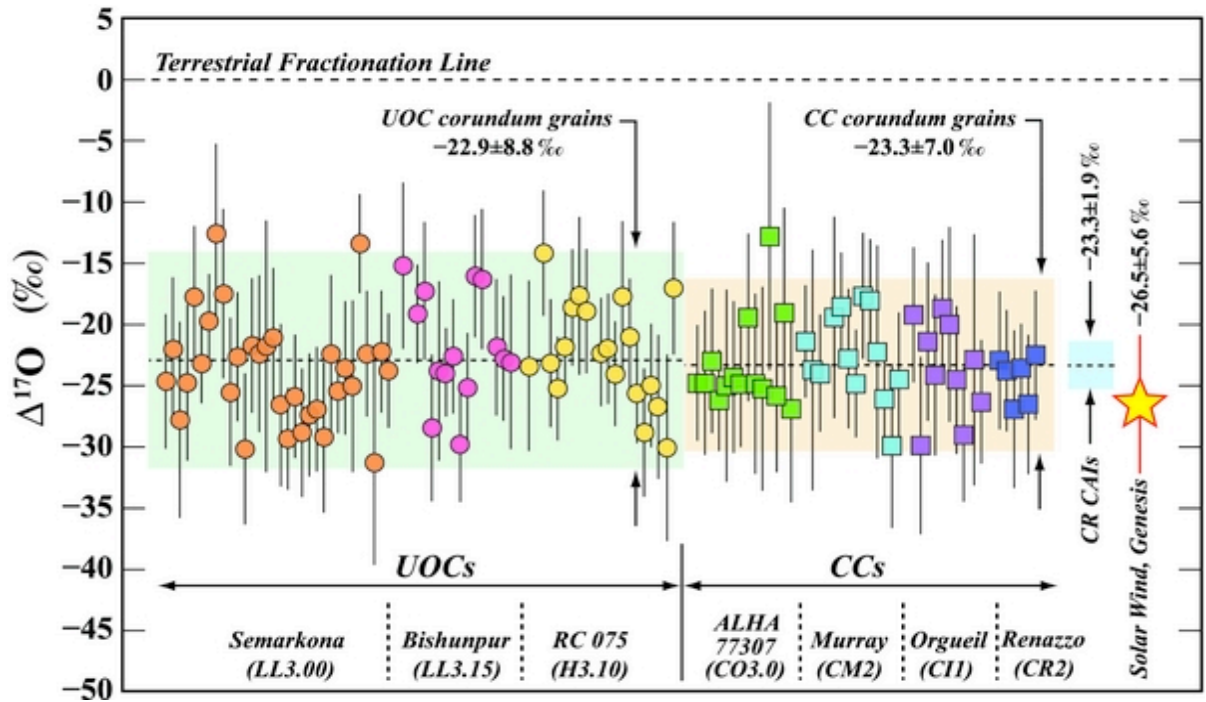

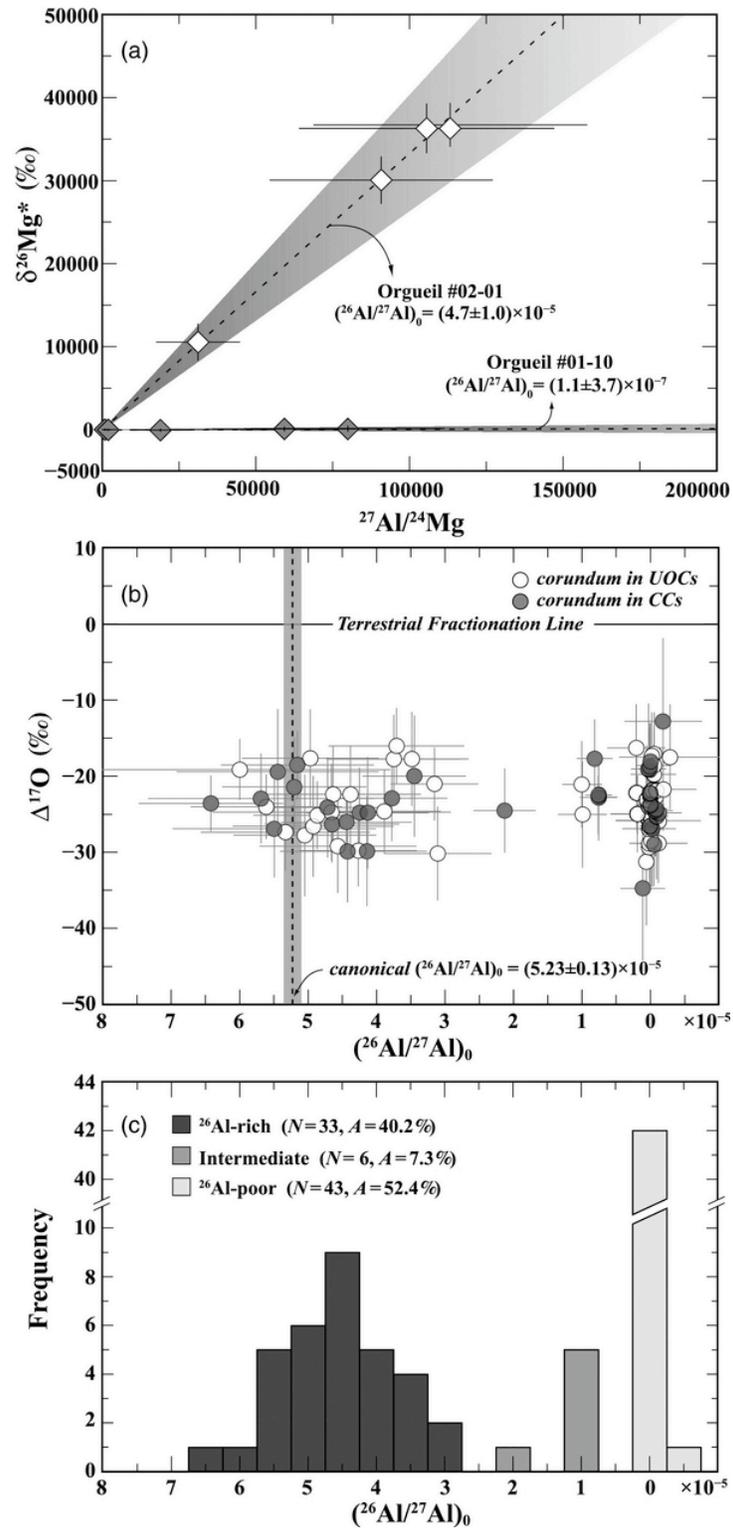

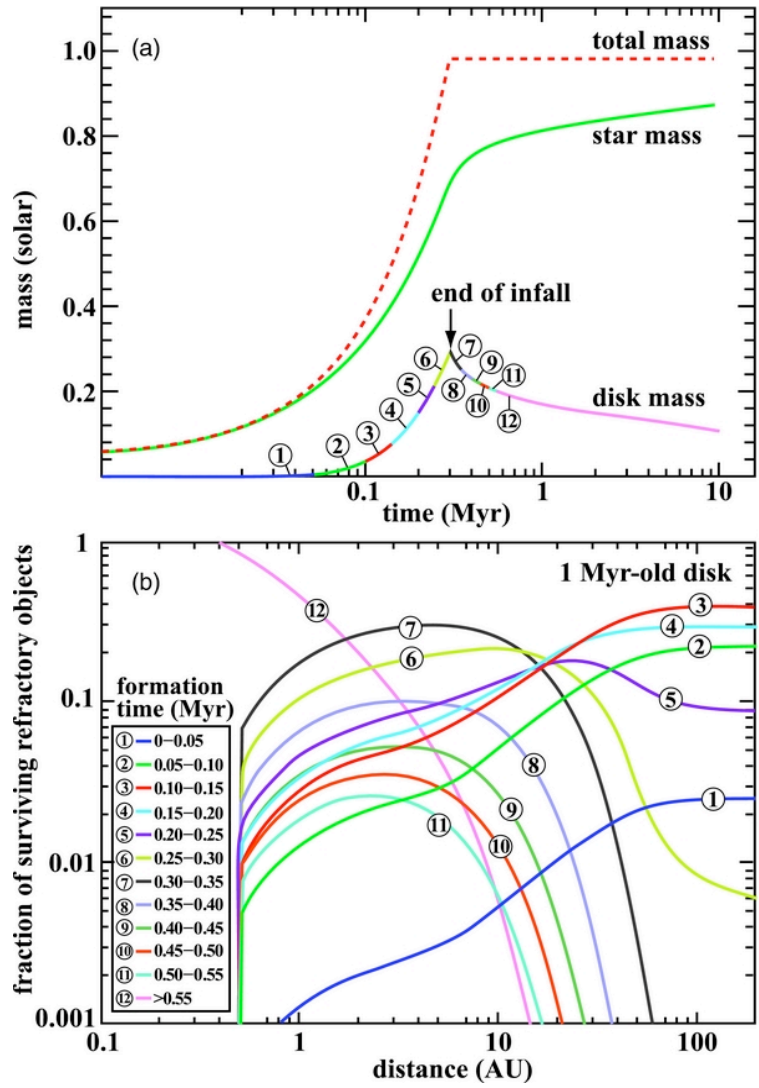